# EMOTION-INSPIRED DEEP STRUCTURE *(EiDS)* FOR EEG TIME SERIES FORECASTING

**Mahboobeh Parsapoor**

## ABSTRACT

Accurate forecasting of an electroencephalogram (EEG) time series is crucial for the correct diagnosis of neurological disorders such as seizures and epilepsy. Since the EEG time series is chaotic, most traditional machine learning algorithms have failed to forecast its next steps accurately. Thus, we suggest a model, which has formed by taking inspiration from the neural structures that underlie feelings (emotional states), to forecast EEG time series. The model, which is referred to as emotion-inspired deep structure (EiDS), can be used to predict both short- and long-term of EEG time series. This paper also compares the performance of EiDS with other variations of long short-term memory (LSTM) networks.

## 1 Introduction

The machine learning (ML) community is interested in developing high-generalization ML algorithms by taking inspiration from cognitive systems. Such ML algorithms can be referred to, variously, as "neuroscience-inspired artificial intelligence" [1], a biologically inspired ML algorithm, a computational intelligence paradigm, or a brain-inspired ML algorithm (i.e., the terminology of this paper). The first step to developing a brain-inspired ML algorithm is to select a cognitive system that has three following criteria (the interested readers may refer to [2] ). The first criterion is that the underlying structure of the cognitive system should encompass several components. The second point is that the cognitive system should fulfill a goal-based (e.g., cognitive) or state-based (e.g., emotional) function and through interaction between its components. The third circumstance is that the cognitive or emotional function should exhibit intelligent behavior (i.e., this means that an intelligent behavior emerges from the results of emotional or cognitive function). Two well-known examples of brain-inspired ML algorithms are convolutional neural networks (CNNs) and long short term memory networks (LSTMs). Even, these models have successfully solved complex classification problems, however they have been developed by taking inspiration from the cognitive systems that accomplish cognitive (goal-based ) functions. Cognitive functions in the brain are fulfilled through iterative processes. Thus, brain-inspired ML algorithms that are based on cognitive functions require many iterations to be able to learn the behavior of data samples. However, ML developers would prefer to solve complex problems using ML algorithms with low computational complexity. Besides, most ML algorithms have complex structures and suffer from high model complexity. These issues are exacerbated by modern, complex, high-dimensional inputs and large-scale problems [3]. Considering these issues, we suggest that a brain-inspired ML algoritham should be designed based on the emotional systems. One of the well-known emotional systems in the brain is the emotional system that is responsible for providing the fear reaction (i.e., the emotional system of fear). Structurally, the emotional system encompasses connected regions of the brain (e.g., amygdala); each region has a specific function, used to process its input, and interacts with other regions to generate its output. Finally, intelligent behavior emerges from the functionality of the emotional system of fear. Thus, the emotional system can be the basis of the development of an ML tool. We are referring it to as emotion-inspired deep structure (EiDs) that is a variation of brain emotional learning-inspired models[2]. The focus of this paper is to demonstrate that EiDS is useful as a time series prediction model to forecast the short- and long-term horizon of the EEG time series.

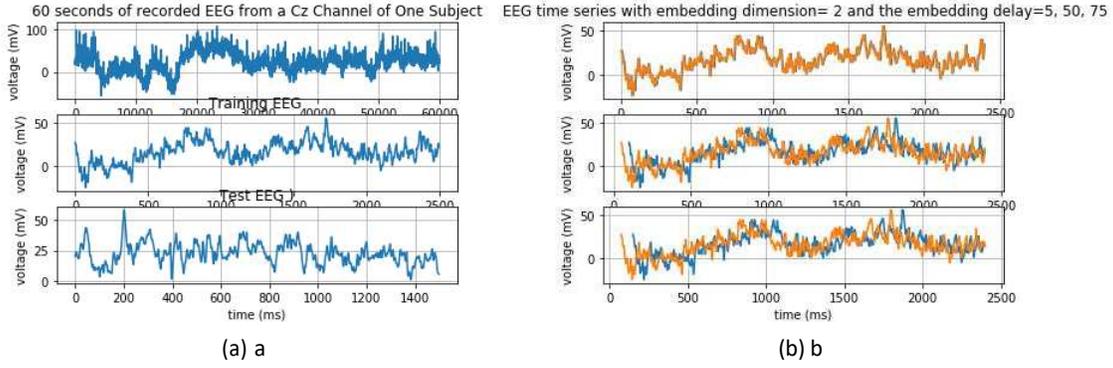

(a) a                                                                 (b) b

Figure 1: (a).Recorded brain activities from one subject, the upper plot is the sixty milliseconds of the EEG values of the Cz channel, the middle and lower plots are the EEG values to construct the training data set and the test data set respectively. (b). Plots of EEG time series with different embedding dimension and embedding lag

## 2   EiDS for Forecasting EEG Time Series

As a time series prediction model, the EiDS receives $(i_{kt-D}, ..., i_{kt-1}, i_{kt})$, and forecasts the value of $i_{kt+1}$. To forecast both long and short horizon of EEG signals, [1], we first construct a univariate EEG time series of the EEG signals of one channel (the Cz channel) of one subject ( see Equation 1).

$$i_j = [EEG(t_j - (D-1)\Delta), ..., EEG(t_j - \Delta), EEG(t_j), EEG(t_j + \Delta)] \tag{1}$$

As a time series prediction model, the EiDS receives $(i_{kt-D}, ..., i_{kt-1}, i_{kt})$, and forecasts the value of $i_{kt+1}$. To forecast both long and short horizon of EEG signals, [2], we first construct a univariate EEG time series of the EEG signals of one channel (the Cz channel) of one subject ( see Equation 1). Here, $i_j$, is the state vector, $EEG(t_j)$ denotes the value of EEG signal at the time, $t_j$, with the prediction horizon, $\Delta$, and the embedding dimension, $D$ [3]. At the second step, we form the training and test dataset. We have considered a set of EEG values collected from 36 to 40 seconds (see Figure 1.(a)). The first 2500 samples of the data set have been considered as the training data set while the next 1400 values have been used as the test data set (see Figures 1.(a)). For short term forecasting of the EEG time series, $[EEG(t_j - 2), ..., EEG(t_j - 1); EEG(t_j)]$, and $[EEG(t_j - 10), ..., EEG(t_j - 5); EEG(t_j)]$, the embedding dimension and the prediction horizon are equal to (2,1ms) and (2,5ms) respectively. For these two examples, as observed from Table 1 the obtained MAEs (mean absolute errors) by the EiDS is equal to 0.85 and 4.25 that are slightly similar to errors of other LSTM networks. For forecasting long term behavior of EEG signals, we examine the EiDS and other LSTM networks to forecast $EEG(t_j)$ by receiving $[EEG(t_j - 100), EEG(t_j - 50)]$, which has been described by the middle curves of Figure 1.(b). For this example, the obtained RMSE (Root Mean Square Error) is equal to 9.72 which is lower than RMSE obtained by bidirectional LSTMs and higher than RMSE of Vanilla and Stacked LSTMs. As another long term prediction example, we have constructed an EEG time series as $[EEG(t_j - 150), ..., EEG(t_j - 75); EEG(t_j)]$ which has been described by the lower curves of Figure 1.(b). According to Table 1 ([4], neither EiDS nor other LSTM networks can accurately predict 75ms ahead of the EEG time series. However, the RMSE of EiDS is lower than other LSTMs.

DL algorithms are highly capable of processing a large amount of data; thus, introducing the EiDS as a new DL tool, we need to investigate its capability to learn from a large data set. As the final experiment of this paper, we have utilized EiDS to forecast the short-term behavior of the EEG time series by using 7000 training samples (see Figure3.(a)). Figure3.(b) describes the convergence curves regarding this example. As the convergence curves [6] show EiDS converges faster than three other LSTMs.

---

[1] EEG signals have been provided by the "Cognitive and Social Neuroscience lab" at the Douglas hospital for further information about data recording and data prepossessing, refer to brain to brain interaction

[2] EEG signals have been provided by the "Cognitive and Social Neuroscience lab" at the Douglas hospital for further information about data recording and data prepossessing, refer to brain to brain interaction

[3] the number of samples of data sets that can be mapped to reconstruct one state vector of the time series; The predictability of a time series depends on its embedding dimension and the prediction horizon

[4] for EiDS, we have three LSTM networks, and each LSTM network can be specified as (No.Hidden Layers and No. Cells)

[6] they show how the training dataset updates the parameters to converge to the minimum value of its loss function

Table 1: Forecasting short and long horizon of EEG data set by BELiM, LSTM

| Specification | | | | | |
|---|---|---|---|---|---|
| Model's Name | Model's Structure | Steps (ms) | RMSE | MAE | No.Iterations |
| Vanilla LSTM | (1,14) | 1 | 0.99 | 0.77 | 1000 |
| Vanilla LSTM | (1,15) | 5 | 5.49 | 4.2 | 1000 |
| Vanilla LSTM | (1,20) | 50 | 9.62 | 7.4 | 2000 |
| Vanilla LSTM | (1,20) | 75 | 10.24 | 8.06 | 1000 |
| Stacked LSTM | (3,9,8,3) | 1 | 0.99 | 0.76 | 1000 |
| Stacked LSTM | (3,15-8-5) | 5 | 5.51 | 4.301 | 500 |
| Stacked LSTM | (3, 15-8-5) | 50 | 9.69 | 7.5 | 500 |
| Stacked LSTM | (3, 20,4,2) | 75 | 10.06 | 7.84 | 500 |
| Bidirectional LSTM | (1,28) | 1 | 1.01 | 0.78 | 1000 |
| Bidirectional LSTM | (1,28) | 5 | 5.47 | 4.27 | 500 |
| Bidirectional LSTM | (1,28) | 50 | 10.28 | 8.11 | 500 |
| Bidirectional LSTM | (1,28) | 75 | 9.79 | 7.6 | 500 |
| EiDS | $((1,6),(1,5),(1,7))_5$ | 1 | 1.10 | 0.85 | (100,150,400) |
| EiDS | ((1,6),(1,5),(1,8)) | 5 | 5.47 | 4.25 | (100,100,100) |
| EiDS | ((1,6),(1,5),(1,8)) | 50 | 9.72 | 7.48 | (100,100,100) |
| EiDS | ((1,6),(1,5),(1,8)) | 75 | 9.63 | 7.4 | (100,100,100) |

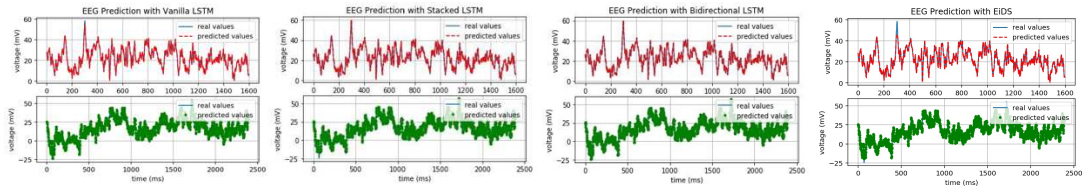

Figure 2: The predicted values versus thee observed values for one millisecond ahead prediction of EEG time series obtained from examining different variations of LSTM networks such as (a).Vanilla LSTM [4], (b).stacked LSTM [5], (c).Bidirectional LSTM and (d).the EiDS. The red curves are predicted values receiving samples from the test data; while, the green lines are predicted values receiving samples from the training data)

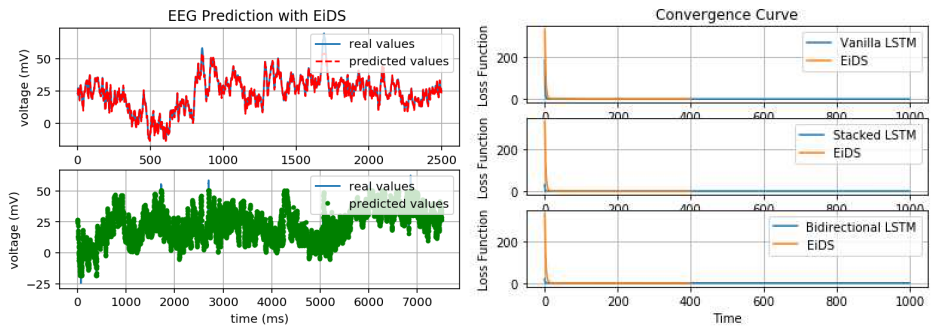

Figure 3: (a).The predicted values versus real values for a large training data sets; (b). Convergence curves of the BELiM and other LSTM networks

## 3  Conclusion

Since accurate forecasting, the long term behavior of chaotic time series is one challenging problem in the ML community. We have developed EiDS to have an ML tool with high performance for long term forecasting of chaotic time series. This paper presented the preliminary results obtained from using EiDS as a time series forecasting model. We also compared EiDS' results with three LSTM networks: Vanilla LSTM, Stacked LSTM, and Bidirectional LSTM networks. The comparison verified that EiDS performs acceptably; however, to have more accurate results and better generalization, we will enhance EiDS. As future work, we want to use the EiDS to quickly and accurately forecast long term horizon of multivariate EEG time series [7]. Such forecasting is essential for our brain-computer interface (BCI) application ( interesting readers can further read about ongoing research projects at Cognitive and Social Neuroscience lab) [6].

**Acknowledgment** The first author is grateful to the "Cognitive and Social Neuroscience lab" for providing access to EEG datasets.

---

[7] 5 millisecond ahead of EEG values of 32 EEG channels